\documentclass[english,prb,floats,superscriptaddress,usenames]{revtex4}
\usepackage[T1]{fontenc}
\usepackage[latin9]{inputenc}
\usepackage{amsmath}
\usepackage{amssymb}
\usepackage{graphicx}
\usepackage{esint}
\usepackage{lipsum}

\usepackage{framed}

\makeatletter

\newcommand{\be}{\begin{equation}} 
\newcommand{\ee}{\end{equation}}
\newcommand{\bea}{\begin{eqnarray}}   
\newcommand{\eea}{\end{eqnarray}}

\usepackage{babel}

\makeatother

\usepackage{babel}
\begin{document}

\date{\today}


\title{Active particles under confinement and effective force generation among surfaces}

\author{Lorenzo Caprini}
\address{Gran Sasso Science Institute (GSSI), Via. F. Crispi 7, 67100 L'Aquila, Italy}

\author{Umberto Marini Bettolo Marconi}
\address{ Scuola di Scienze e Tecnologie, 
Universit\`a di Camerino, Via Madonna delle Carceri, 62032, Camerino, INFN Perugia, Italy}


\date{\today}							

\begin{abstract}

 We consider the effect of geometric confinement 
 on the steady-state properties of a one-dimensional active suspension subject to thermal noise. The random active force is modeled by an
Ornstein-Uhlenbeck process and the system is studied both numerically, by integrating the Langevin governing equations, and analytically
by solving the associated Fokker-Planck equation under suitable approximations. The comparison between the two approaches displays a fairly good agreement and in particular, we show that the Fokker-Planck approach can predict the structure of the system both
in the wall region and in the bulk-like region where the surface forces are negligible.
The simultaneous presence of thermal noise  and active forces determines the formation of a layer,
extending from the walls towards the bulk, where the system exhibits polar order.
We relate the presence of such ordering to 
 the mechanical pressure exerted on the container's walls and show
how it depends on the separation of the boundaries and determines a Casimir-like attractive force mediated by the active suspension.
\end{abstract}
\maketitle

\section{Introduction}
Self-propelled particles, motile organisms such as bacteria, and artificial micro-swimmers display a characteristic tendency to aggregate, a phenomenon which is currently the object of vivid interest among physicists and biologists \cite{ramaswamy2010mechanics, romanczuk2012active,marchetti2013hydrodynamics,saintillan2015theory}. Unlike standard molecular systems, where aggregation is induced by attractive forces and/or entropic interactions \cite{likos2001effective} due to volume exclusion, active particles may spontaneously produce
regions of higher density because their dynamical properties change if they interact with other particles.
These phenomena have been investigated experimentally, by numerical simulation and theoretically \cite{bechinger2016active,stenhammar2014phase}
and led in the case of self-propelled particles endowed with only repulsive inter-particle interactions to the concept
of motility induced phase separation (MIPS) \cite{cates2015motility} analogous to the liquid-gas coexistence in standard liquids. On the other hand,
the accumulation phenomenon in the proximity of a purely repulsive confining wall, i.e. the aggregation with an external object, occurs even when active particles are not subject to mutual interactions \cite{lee2013active}. Such a behavior is of great practical importance since experiments are often conducted on systems where the size of the experimental apparatus could be of an order of magnitude comparable to the persistence length of the active particles, which is the typical distance over which particle's orientation persists. The explanation of the underlying mechanism attributes the accumulation to the reduction of the particles' mobility in the presence of the walls and is captured by some existing theories \cite{ezhilan2015distribution,maggi2015multidimensional,nikola2016active,fily2014dynamics,smallenburg2015swim,marconi2017self}.

Confining surfaces besides triggering particle accumulation in a very thin adjacent region, may also create a diffuse layer
where neither the density is constant nor polar order field vanishes as in bulk systems.
In a series of recent articles Brady and coworkers~\cite{yan2015force,yan2018curved} have thoroughly investigated
such an inhomogeneous layer by means of a mesoscopic approach where these effects
were captured by a simple system of differential equations and
the action of the walls was taken into account by prescribing the appropriate boundary conditions.

In the present article, we consider
the effect of a confining potential, $\phi(x)$, varying along the single $x$-direction, on the steady-state behavior of an assembly of non-interacting self-propelled particles described by means of the so-called  active Ornstein-Uhlenbeck particle (AOUP) model \cite{fily2012athermal,szamel2014self} 
 The AOUP is driven by an active force of variable intensity and direction assimilated to a Gaussian colored noise
process \cite{hanggi1995colored} sharing the same exponential two-time correlator as the active force of the ABP model.

The characteristic time, $\tau$, of the process represents the average persistence of the trajectory along a given spatial direction, i.e. the crossover time from a ballistic to a diffusive behavior.  In fact, in both models the mean square displacement
evolves ballistically at short time and diffusively at long times with an effective diffusion coefficient given by the sum of 
an active contribution,  $D_a$  plus a thermal contribution,  $D_t $, stemming from the microscopic collisions with the solvent molecules.
Regarding the difference between the two models, in ABP the absolute value of the active speed is constant, whereas in AOUP
each component independently fluctuates according to a Gaussian distribution.
Our choice to use the AOUP, instead of other popular alternatives such as the ABP and the Run and Tumble~\cite{tailleur2008statistical} models is motivated by the possibility of applying straightaway methods and concepts similar to those employed in the study of the Kramers equation \cite{titulaer1978systematic}.
In the last few years, approximate treatments of the AOUP, such as the Fox method of Ref.~\cite{farage2015effective,wittmann2017effective} and the so-called unified colored noise approximation (UCNA)~\cite{marconi2015towards}, have been developed:
by using an adiabatic approximation, which is tantamount to impose a detailed balance condition~\cite{cates2012diffusive}, the UCNA allowed making reasonably accurate predictions about the steady-state properties of a rather general class of active systems  \cite{marconi2016pressure,marini2017pressure}.
Nevertheless, the UCNA results regarding the structure of active suspensions
in the proximity of a confining surface disagree with those obtained by mesoscopic treatments of the ABP model, where the container wall is treated as an infinitely sharp interface and a set of boundary conditions on the density and polar fields are imposed on it.
Such a discrepancy is more severe when the finite diffusivity of the solvent, $D_t$, is not negligible and a polar order appears close to the 
surface.
In the present theory, we go beyond the UCNA and do not impose the detailed balance condition in deriving the form of the steady-state
solution. In contrast with mesoscopic approaches ~\cite{ezhilan2015distribution,yan2015force,duzgun2018active}, we treat the wall and bulk regions on equal footing
and instead of considering the wall as a sharp boundary~\cite{fily2017equilibrium} we study the distribution function in each region thus providing a microscopic description of
why and how particles accumulate at the boundaries and form a diffuse layer near it.

 At variance with the ABP model which is well defined only for two or more dimensions, the AOUP model can be implemented also in one-dimension. In a system with a simple geometry, such as infinite parallel plates - a situation which can be realized assuming periodic boundary conditions - the coordinates parallel to the plate play just a minor role, as a constant factor in the definition of the control parameters. In practice,  provided we restrict to a region far enough from the edges of the plates
 the present treatment applies also to the case of finite plates and the one-dimensional description is valid.

Besides clarifying the mechanism causing the enhancement of  the density and polar order near the walls in self-propelled systems,
we discuss the role of the activity in determining the forces that the particles exert on inclusions, a topic of current interest.
In fact, several groups by numerical simulation of  RnT~\cite{ray2014casimir}, ABP~\cite{ni2015tunable} and swimmer suspensions \cite{parra2014casimir}, have recently reported evidence about effective interactions
arising between two plates placed in a bath of active particles, a phenomenon akin to the Casimir-like attractive force~\cite{casimir1948attraction} observed in the presence of non-equilibrium diffusive dynamics~\cite{aminov2015fluctuation,brito2007generalized}.

The paper is organized as follows: in Section \ref{Model}, we introduce the model of confined active particles
and consider a truncated parabolic confining potential~\cite{solon2015pressure,sandford2017pressure}.
In Section~\ref{Numericalsimulation}, we illustrate the numerical method and study the model by numerical simulations,
in Section~\ref{Theory} we present our theory
which goes beyond the UCNA concerning an important aspect: for a system of non-interacting particles the UCNA predicts that the
distribution function has a local dependence on the potential, thus if the potential and its derivatives vanish in some region of space
the distribution is uniform. By using a hierarchy of equations for the velocity-moments of the phase-space distribution we are able
to describe non-local effects and obtain predictions which are in better agreement with the numerical simulation results.
In section \ref{Pressure}, we discuss the pressure exerted on the walls by the active suspension
 using the results of Section~\ref{Theory}. As an application we derive a new expression for the effective force between two plates
induced by the activity when the molecular diffusion is finite. Finally in Sec.~\ref{Conclusions} we present the conclusions.

\section{Model}
\label{Model}

The model consists of a system of 
 $N$ non-interacting active particles suspended in a fluid,
 driven by an active force $\gamma u$, where $\gamma$ is a  Stokes friction constant, and subjected to an external potential $\phi(x)$ and to a random
 force representing the effect of the collisions with the molecules of the fluid. 
  The self-propulsion force, originating from an internal mechanism and fluctuates both in intensity and direction and is modeled by a
 colored noise term, $u(t)$, evolving according to an Ornstein-Uhlenbeck process
of correlation time, $\tau$. The resulting governing equations read:
\bea
&&\gamma\dot{x}=-\phi'(x) + \gamma u+\gamma\sqrt{2D_t}\xi, \label{dotx}\\
&&\tau\dot{u}=-u+\sqrt{2D_a } \eta,
\label{dotu}
\eea
where $\xi$ and $\eta$ are two independent white noises with unitary variance and zero average and $(D_t,\gamma)$ and $(D_a,\tau)$ refer to the interactions with the solvent and with the active bath, respectively. 
The term $\gamma u$
represents the self-propulsion mechanism, an
internal degree of freedom converting energy into motion and has the following self-correlation $\langle u(t) u(t')\rangle=\frac{D_a}{\tau} \exp( -|t-t'|/\tau)$, with variance 
$D_a/\tau$  identified with the active power. 
It is well-known that this system is out of equilibrium whenever $\tau>0$~\cite{marconi2017heat} and
 that in the limit $\dot{u} \approx 0$  Eq.~\eqref{dotu} reduces to
$u \approx \sqrt{2D_a}\eta$, a Wiener process so that Eq.~\eqref{dotx} describes a Brownian passive particle where the term $\gamma u$ merely produces an extra contribution to the diffusion.

We, now, generalize to the case $D_t\neq 0$ the change of variable of ref~\cite{marconi2016velocity} which allows a hydrodynamic study
of the model: we define the new variable  $v=\dot{x}-\sqrt{2D_t}\xi$ and
replace Eqs.~\eqref{dotx} and \eqref{dotu} by the following set of equations:
\begin{flalign}
&\dot x=v +\sqrt{ 2 D_t}  \xi
\label{sdotx}
\\
&\dot v=-\frac{1}{\tau} \Bigl[ v+\frac{\phi'(x)}{\gamma} -\sqrt{ 2 D_a}  \eta\Bigr] -\frac{\phi''(x)}{ \gamma}  v - \frac{\sqrt{ 2 D_t} }{\gamma} \phi''(x)\xi.
\label{dotv}
\end{flalign}
Thus the AOUP dynamics Eqs.~\eqref{dotx}-Eq.~\eqref{dotu}  has been mapped onto the underdamped dynamics of a fictitious Brownian particle of position $x$ and velocity $v$ and effective mass $\mu=\tau \gamma$ evolving with a space dependent Stokes force and experiencing
 a delta-correlated thermal noise acting additively on the $x$ and multiplicatively on the velocity~\cite{caprini2018linear}. Given the presence of multiplicative noise terms we use the Stratonovich
interpretation of the stochastic differential equation~\cite{hanggi1995colored}.

For mathematical convenience, we shall restrict our study to the case where the effect of the confining walls is represented
by two repulsive truncated harmonic potentials, $\theta(-x)k x^2/2$ and $\theta(x-2L)k (x-2L)^2/2$  ($\theta$ being the Heaviside function)
acting only in the regions $(-\infty,0)$ and $(2L, \infty)$, whereas the central  region ($0,2L)$ is a force-free region.
 The harmonic force is proportional to  $k$, modeling the penetrability of the wall: since the 
range and strength of the force are both finite it could describe an elastic membrane of stiffness $k$ allowing the particles to explore the regions $x\leq 0$ and $x\geq 2L$. 
 On the other hand, if the spatial resolution of the experimental device is low or the penetrability of the wall is small  
the use of a sharp interface model, obtained by imposing  no-flux boundary conditions to prevent particle crossings as in ref.~\cite{yan2015force},  is well justified. 

Finally,
with the aim of developing the theoretical methods of Sec.~\ref{Theory}
 we introduce the stationary Fokker-Planck equation (FPE)~\cite{risken}  for the phase-space distribution $f(x,v)$
 providing an equivalent statistical description of the system \eqref{sdotx}-\eqref{dotv}. We 
 substitute $\phi'(x)= k [x\theta(-x)+(x-2L)\theta(x-2L)] $ and $\phi''(x)= k [\theta(-x)+\theta(x-2L)] $
and obtain the following equation:
\begin{equation}
\begin{aligned}
&  v \frac{\partial f(x,v)}{\partial x}-D_t \frac{\partial^2 }{\partial x^2} f(x,v) \\
& -  \frac{1}{\tau} \frac{\partial }{\partial v} \left(\frac{D_a}{\tau} \frac{\partial }{\partial v}+\left(1+
\left( \theta(-x)  + \theta(x-2L) \right) \frac{\tau k}{\gamma} \right) v   \right) f(x,v)
=\\
&\left( \theta(-x) \frac{kx}{\tau\gamma} + \theta(x-2L) \frac{k(x-2L)} {\tau\gamma}\right) \frac{\partial f(x,v)}{\partial v}
\\
&+ \left( \theta(-x)  + \theta(x-2L) \right) \frac{D_t}{\gamma} k \left(  \frac{k }{\gamma}  \frac{\partial^2 }{\partial v^2}f(x,v)
- 2\frac{\partial^2 }{\partial v \partial x }   \, f(x,v)  \right) ,
\end{aligned}
\label{fpeq}
\end{equation}
where $\gamma/k$ is the characteristic time of the $x$-process.


\section{Numerical methods and results}
 \label{Numericalsimulation}

  In our numerical simulations
 equations~\eqref{dotx}-\eqref{dotu} have been integrated by using the Euler-Maruyama algorithm neglecting terms of order $dt^{5/2}$, where $dt$ is the time-step size of the numerical integration \cite{mannella1989fast}.
Except where noted, each simulation has been run until time $2\cdot10^3\tau$, with $dt\sim O(10^{-4} - 10^{-6})$, depending on the $\tau$ values.
 The observables, such as the probability distribution and its momenta, have been computed by using both time and ensemble averages: we usually perform simulations with $N=10^4$ particles, waiting for a transient time $10^3\tau$, in such a way the system is fully thermalized. 
 
\begin{figure}[!t]
\centering
\includegraphics[width=1\linewidth,keepaspectratio]{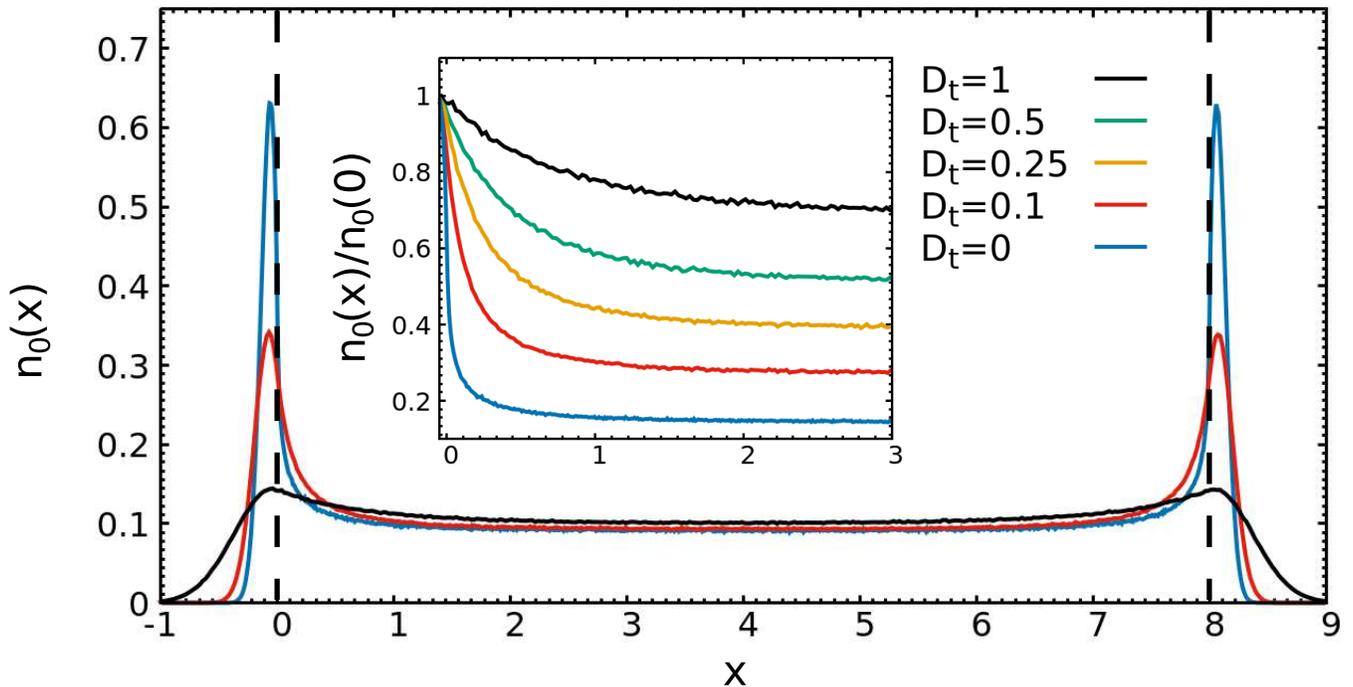}
\caption{Density profile, $n_0(x)$, for a system with symmetric harmonic walls 
placed at a distance $2L=8$ at $x=0$ and $x=8$. The curves correspond to simulation results for different values of $D_t$ as shown in the legend, while $D_a=1$, $\tau=1$ and $k=10$ are kept fixed. In the inset:  rescaled density profiles, $n_0(x)/n_0(0)$.  }
\label{fig:graphuno}
\end{figure}

In Fig.~\ref{fig:graphuno} we display the density profile obtained by numerical simulation in the case of two walls separated by a distance $2L=8$ keeping constant the ratio $D_a=1$ and $\tau=1$ and varying the intensity of the thermal noise $D_t$
as shown in the legend. One can observe that the density profile, $n_0(x)$, is continuous for all values of $D_t$ including the value $D_t=0$,
at variance with the UCNA which predicts a finite jump at $x=0$ and $x=2L$~\cite{fily2017equilibrium}.
The effect of increasing the thermal diffusion, $D_t$,  is to broaden the distribution with respect to the case $D_t=0$ and
is best appreciated in the inset which shows that the profiles 
corresponding to the larger values of $D_t$ have slower decay.
Such a scenario is similar to the one observed in the ABP model \cite{yan2015force}, where the presence of thermal noise has two consequences: a) it reduces the accumulation near the walls and b)
 it determines an exponential decay of the density profile in the force-free region and an associated screening  length, $\lambda$, roughly dependent on the ratio $\Delta=D_t/D_a$.
It is also interesting to see that the accumulation phenomenon near a repulsive wall, a typical non-equilibrium effect,
survives upon the addition of thermal noise and disappears only in
 the limit $D_t\gg D_a$ when the particles behave as passive ones.
 We may conclude that $D_t$ has a double role in the potential region: on one hand, reduces the accumulation, decreasing the height of the peak; on the other hand, it favors the dispersion for $x>0$ as shown in the inset  of Fig.~\ref{fig:graphuno}.

\begin{figure}[!t]
\centering
\includegraphics[width=1\linewidth,keepaspectratio]{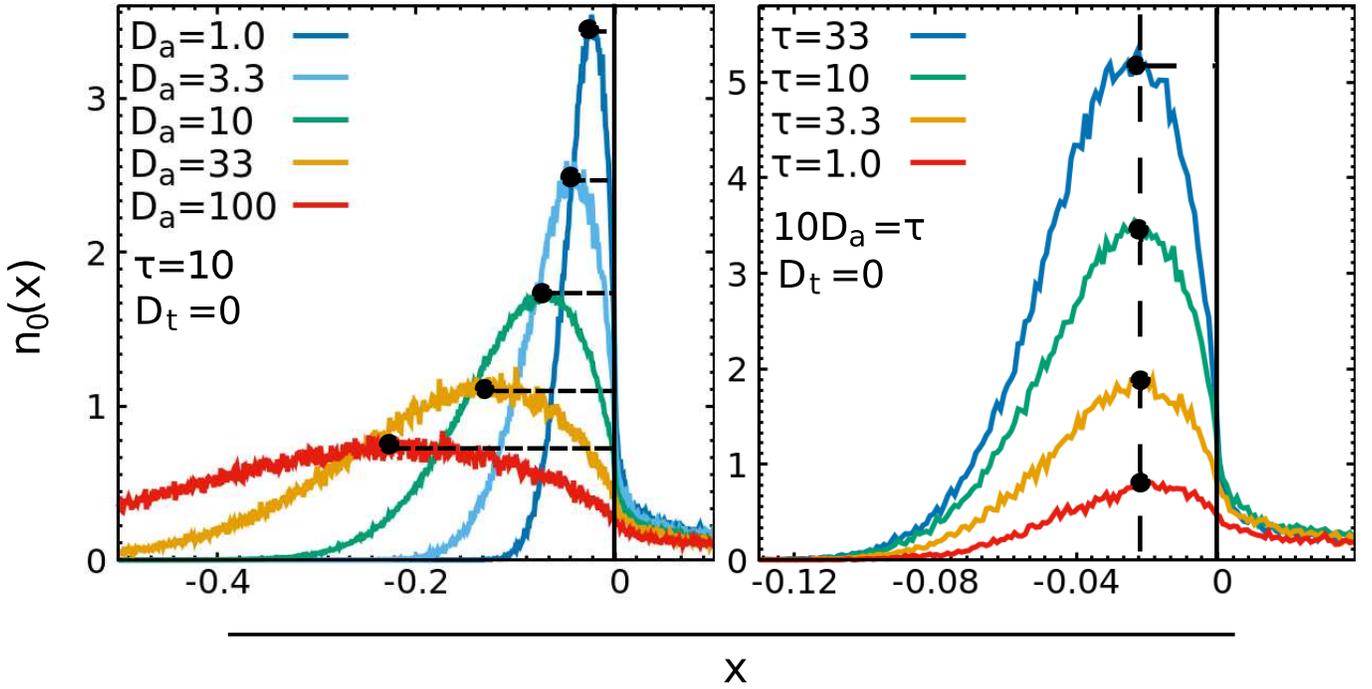}
\caption{ Density profiles, $n_0(x)$, for a system with $D_t=0$, $k=10$ and symmetric harmonic walls 
placed at a distance $2L=8$  at $x=0$ and $x=8$. 
In these panels, for presentation purpose, we only show $n_0(x)$ in the proximity of the left wall. Left panel:  each line corresponds to a system with $\tau=10$ fixed and different values of $D_a$ in an interval between $1$ and $100$,  (and so varying the active power $D_a/\tau$). Notice that the peak broadens and shifts towards more negative values of the x-coordinate
with increasing active power.
Right panel: Simulation results with $D_t=0$  and  different values of $\tau$, keeping constant the ratio $D_a/\tau$.}
\label{fig:graphunodoppio}
\end{figure}

In Fig.~\ref{fig:graphunodoppio} we analyze the system with $D_t=0$  discussing for the sake of simplicity  just the left wall:
the particles accumulate approximately in the region close to $x=0$ and their profile does not have a Brownian counterpart: in fact, in the Brownian limit  $\tau \to 0$, they would be uniformly distributed between $0$ and $2L$ and depleted 
within the repulsive regions according to the Boltzmann weight $\propto \exp(-\phi(x)/T)$ at a uniform temperature $T=(D_a+D_t)\gamma$.
If $\tau>0$, the accumulation can be understood by considering the Eq.\eqref{fpeq}  (with $D_t=0$): the Stokes force is discontinuous, being $\gamma \Gamma v$ for $x<0$, 
with $\Gamma=1+\tau k /\gamma$,  and $\gamma v$ for $x>0$. Hence, on one hand particles slow down in the regions $[-\infty, 0]$  and on the other hand the repulsive wall pushes the particles towards the edge $x=0$. The interplay between the slow-down and the repulsion determines the observed accumulation.
 In the left panel of Fig.~\ref{fig:graphunodoppio} each line corresponds to a system with $\tau=10$  and $D_a$ ranging between $1$ and $100$. 
 Notice that the peak broadens and shifts towards more negative values of the x-coordinate with increasing active power $D_a/\tau$.
 The right panel  of Fig.~\ref{fig:graphunodoppio} shows that the location of the peak does not change if $D_a/\tau$ remains constant but its height
 increases when  $\tau$ increases: indeed, a larger  $\tau$ corresponds to a longer time spent by the particle in front of the wall and has no influence on the peak dispersion.

 \section{Theoretical treatment}
 \label{Theory}
 
 Now, we present a theoretical analysis of the confined active system using a velocity-moment expansion to derive approximate solutions
 of the FPE~\eqref{fpeq} and compare the theoretical predictions with the numerical solutions of Eqs. \eqref{dotx}-\eqref{dotu}. To this purpose
 it is mathematically convenient  to study separately the two boundary regions characterized by a finite value of the external field from the 
 central force-free region and write the stationary distribution associated with the FPE~\eqref{fpeq}  as
$f(x,v) =\theta(-x)f_l(x,v) + \theta(x)\theta(L-x)f_c(x,v) + \theta(x-L) f_r(x,v)$,
where $f_l$ and $f_r$ are the distribution functions in the left and right regions, respectively, while $f_c$ is the distribution in the central region.

\subsection{Density profile in the wall region}
\begin{figure}[!t]
\centering
\includegraphics[width=1\linewidth,keepaspectratio]{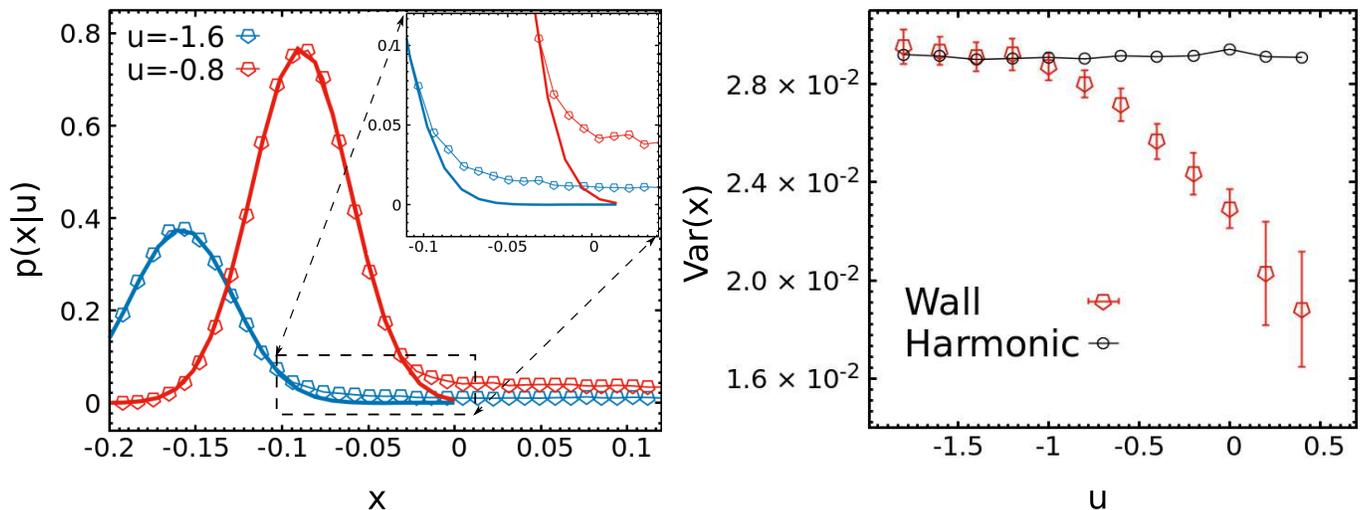}
\caption{Left panel:  Comparison between 
the conditional probability distribution $p(x|u)$ (data) and a fitted Gaussian (solid line), for two different negative values of $u$. Right panel: 
We show for different values of $u$ the comparison between the $x$-variance, $\int dx x^2 p(x|u)$  obtained from numerical simulations (red symbols) and the corresponding variance computed using the 
harmonic potential formula \eqref{pxuharm}. Parameters: $k=10$, $D_a=\tau=1$, $D_t=0$.}\label{fig:pdf_diff}
\end{figure}

 In order to derive a theoretical expression for the probability density $n_0(x)$
in agreement with the numerical results above illustrated, we first consider the region $x<0$ and
leave the treatment of $n_0(x)$ in the central region, $0\leq x\leq L$, until Sec.~\ref{centralregion}.
Neglecting the truncated shape at $x=0$ of the potential 
$\frac{k x^2}{2} \theta(-x)$ and using the $(x,u)$ representation,
we begin by approximating the
probability  distribution $p_l(x,u)$ by the stationary distribution, $p_h(x,u)$ of an AOUP confined to a symmetric harmonic trap $(\frac{k x^2}{2} )$ given by ~\cite{das2018confined}: 
\begin{equation}
p_h(x,u)={\cal N} e^{- \left(\frac{ k x^2}{2  D_a\gamma}  \frac{1} {\frac{1}{\Gamma}+\Delta }    \right)}
\, e^{-\left(\frac{\tau\Gamma }{2 D_a }\, \left( u- \frac{k}{\gamma} \frac{x}{1+\Delta} \Gamma \right)^2 \right)} ,
\label{pxuharm}
\end{equation}
${\cal N}$ being a normalization factor and $\Delta=D_t/D_a$.
A similar expression for $f_h(x,v)$ is reported in Eq.~\eqref{statpxv}  if we employ the $(x,v)$ representation.
The reduced probability distribution $n_0(x)= \int du p_h(x,u)$ 
 computed using formula~\eqref{pxuharm}  shows 
 a poor agreement with the numerical results of Fig.~\ref{fig:graphunodoppio},
 since it is a Gaussian centered at $x=0$, in contrast with  the numerical evidence illustrated in Figs. \ref{fig:graphuno} and \ref{fig:graphunodoppio}
where such a peak is shifted towards negative x-values.
 The domain of validity of the harmonic approximation~\eqref{pxuharm} can also  be tested by comparing
the simulation results for the conditional probability distribution function, $p(x|u)=p(x,u)/p(u)$ (where $p(u)=\int dx p(x,u)$) with the corresponding quantity obtained from the theoretical expression Eq.~\eqref{pxuharm}. In the left panel of Fig.~\ref{fig:pdf_diff},
one observes significant deviations between the two curves when $x>0$.

Instead, in the right panel of  Fig~\ref{fig:pdf_diff}, we display the comparison between the 
theoretical (computed from the Gaussian
formula \eqref{pxuharm}) and numerical
 x-variance, $\int dx x^2 p(x|u)$: one can see that the first decreases as $u$ increases, whereas 
 the latter remains nearly constant. The departure from the Gaussian prediction, becomes 
 more and more relevant when $u$ increases, while it is negligible for negative values of $u$.


\begin{figure}[!t]
\centering
\includegraphics[width=1\linewidth,keepaspectratio]{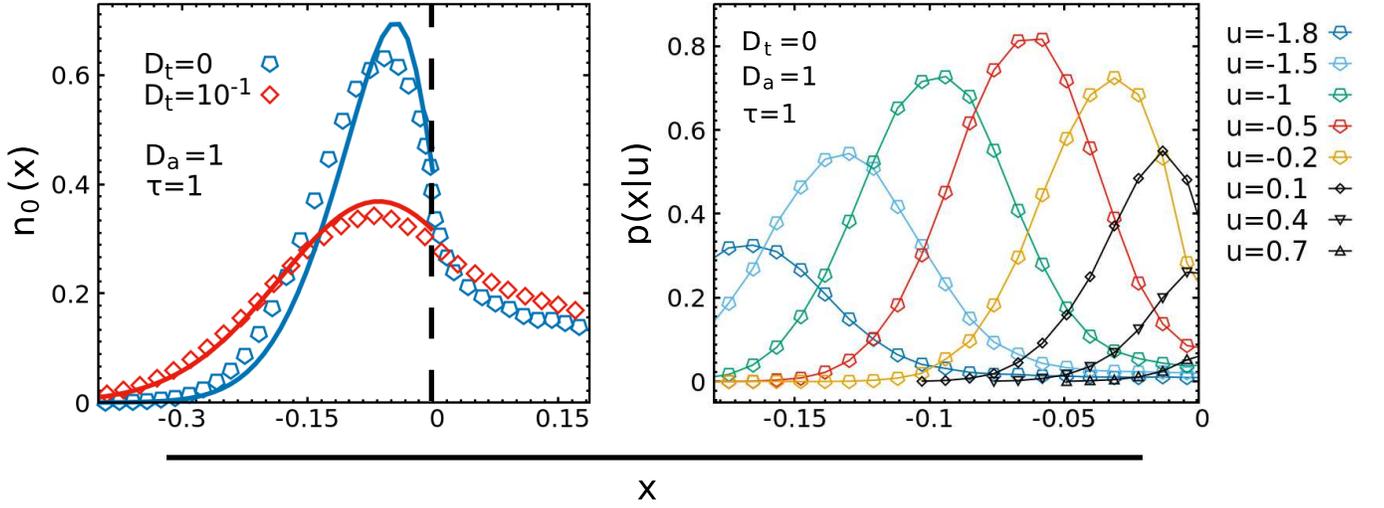}
\caption{Left Panel: Density profile, $n_0(x)$, in the wall region for $D_t=0, 10^{-1}$, respectively blue and red curve. 
Circles represent the numerical data and lines the theoretical prediction Eq.~\eqref{n0error}. Right Panel:
Conditional probability distribution  $p(x|u)$ as a function of $x$ for different values of $u$ as reported in the legend: the colored curves represent the data with $u\leq 0$ and the three black curves the data for $u\geq 0$.Parameters: $D_a=1$, $\tau=1$, $k=10$.}
\label{fig:XatU} 
\end{figure}

 The comparisons shown in Fig. \ref{fig:pdf_diff} , indicate 
that the Gaussian formula~\eqref{pxuharm} agrees quite reasonably with the numerical results
for the conditional probabilities $p(x|u)$ and $p(u|x)$, when $u<0$ and $x<0$, respectively, but 
the same formula does not provide an adequate 
prediction for the density profile $n_0(x)$ for $x<0$.
Based on these evidences we improve the Gaussian approximation~\eqref{pxuharm} 
by modifying the left wall $(x,u)$ distribution in the following way:
\begin{equation}
p_l(x,u) \approx p_h(x,u) \theta(-u) \theta(-x)
\label{kincondition}
\end{equation}   
and an analogous expression for right wall distribution, i.e. $p_r(x,u) \approx p_h(x,u) \theta(u) \theta(x-L)$.
The rationale for such an assumption is the following remark:  the wall can be regarded as
a very massive body having zero speed and the active particle as a moving object with  self-propulsion force $\gamma u$.
A collision between the left wall and the active particle occurs only if $u$ assumes negative values
and $x<0$. These two conditions are encapsulated in formula~\eqref{kincondition}
and we, now,  use it to derive an expression for $n_0(x)$ in the wall region $x \leq 0$ by integrating with respect to $u$. The resulting
density, $n_0(x)$, in the region $x\leq 0$, reads:
\begin{equation}
n_0^{left}(x)=
n_w e^{- \left(\frac{ k x^2}{2  D_a\gamma}   \frac{1} {\frac{1}{\Gamma}+\Delta }     \right) }
\text{erfc}\left(  \sqrt{\frac{\tau \Gamma}{2D_a}}\frac{1}{(1+\Gamma \Delta) } \frac{k}{\gamma}x \right)\theta(-x) ,
\label{n0error}
\end{equation}   
where 
 $n_w$ is the density at $x=0$ and $x=2L$.
Formula~\eqref{n0error} shows a fairly good agreement with the simulation data both for $D_t=0$ and for $D_t>0$, as the left panel of Fig.~\ref{fig:XatU} reveals.

 Eq. \eqref{n0error}, which describes the space-density in the potential regions, generalizes the result of 
\cite{yan2015force}  to a soft-wall, modeled as an external truncated potential. Moreover, we overcome the unphysical results of the UCNA approximation at $D_t=0$  and for hard walls, i.e. a discontinuous space density discussed in ref.\cite{fily2017equilibrium}. 

In order to assess the ansatz $p_l(x,u) =p_h(x,u) \theta(-u)\theta(-x)$ , we analyzed the numerical joint probability distribution $p(x|u)$, at fixed $u$  
and estimated how important is the neglected contribution due to the population with $u\geq0$.  The right panel of
Fig.~\ref{fig:XatU} shows  that in the region $x\leq 0$ the population characterized by a positive sign of the active force $p(x,u>0)$ represents only a small contribution to the density $n_0(x)$ for $x\leq0$, thus roughly validating the approximation leading to Formula~\eqref{n0error}. One can observe that the  $p(x,u\geq0)$ decreases much faster than $p(x,u\leq 0)$  as $x$ becomes more negative. Interestingly, a similar scenario was reported by Widder and Titulaer for a related model \cite{widder1989moment}: these authors studied the distribution function in the presence of a partially absorbing wall with specular reflection
and found that  $p(x,u)$  at the boundary $x=0$ was peaked at negative values of $u$ and rapidly decreasing towards zero
for positive $u$.

  Let us remark that the argument of the complementary error function in Eq.~\eqref{n0error} is proportional to the ratio between the wall force $-kx$ and the average absolute value of the active force, $\gamma \sqrt{D_a/\tau}$.   On the other hand, if $\tau=0$ there is no shift and as we shall see below the accumulation phenomenon is completely suppressed and 
on the contrary, one observes a depletion of the density controlled by the standard Maxwell-Boltzmann weight.
It is possible to define an effective potential  as
$U_{eff}(x)=-\ln n_0^{left}(x)$ and obtain:
\begin{flalign}
U_{eff}(x)&=\frac{k}{ D_a \gamma}\frac{1}{(\frac{1}{\Gamma}+ \Delta)}  \frac{x^2}{2} - \ln   \left( 1+  erf \left(  \sqrt{\frac{\tau }{2 D_a \Gamma }} \frac{k}{\gamma} \frac{|x|}{(\frac{1}{\Gamma}+ \Delta)}  \right)  \right) \nonumber\\
&\approx \frac{1}{(\frac{1}{\Gamma}+ \Delta)} \left( \frac{k }{2 D_a \gamma}\, x^2 -\frac{2}{\sqrt{\pi}}   \sqrt{\frac{\tau }{2\Gamma D_a }} \frac{k}{\gamma} |x|\right) .\nonumber
\label{ueffwall}
\end{flalign}
For small $|x|$ and not too stiff walls ($k\gg 1$ and/or $D_a/\tau\ll 1$) we find that the effective force vanishes when
$$
 x_p = -   \sqrt{ \frac{2}{\pi  \Gamma} D_a  \tau} , 
$$
whereas for strong walls $x_p\approx 0$.
The position of the peak of the distribution does not depend on the ratio $\Delta$ of the two diffusion coefficients,
but its width does. The position, $x_p$, of the peak gives a measure of the stiffness of the wall
and we expect that for $k$ large enough the wall is quite impenetrable and $x_p\approx 0$, while for smaller values of $k$ we have $x_p<0$,
a situation describing a soft wall or a floppy membrane \cite{rodenburg2017van}.

\subsection{Central region}
\label{centralregion}
As illustrated in the previous section, the predictions for the density profile relative to the potential regions are in good agreement with the numerical data. Nevertheless, in the force-free regions (both for $D_t=0$ and $D_t\neq0$) 
one observes a phenomenology which cannot be captured by simply setting to zero the external potential in Eq.~\eqref{pxuharm}, 
with the result of producing a constant $n_0(x)$. In particular, the numerical $n_0(x)$ displays a smooth decay from the wall value towards the value at midpoint $x=L$. Hereafter, we develop a hydrodynamic approach in order to find an approximation scheme for $n_0(x)$ and for this reason we consider appropriate to switch again to the $(x,v)$ representation of the distribution function.
Since for a uniform system  $f(x,v)$  is  a  Maxwell-Boltzmann distribution, in  the force-free region we may expect to find a good approximation by expanding $f(x,v)$  in Hermite functions of the velocity and taking into account only the first few terms.
To this purpose, we seek for an approximate solution of the FPE~\eqref{fpeq} in the central region by employing the following Hermite expansion:
 \begin{equation}
 f_c(x,v)=(\frac{\tau}{2 \pi D_a})^{1/2}\sum_{\nu \geq 0} n_\nu(x) h_\nu(v)\, \exp(- \frac{\tau}{2 D_a} v^2) ,
 \label{hermiteexpansion}
 \end{equation}
 where the Hermite polynomials are
 \begin{equation}
  h_{\nu}(v)
= (-1)^{\nu} (\frac{D_a}{\tau})^{\nu/2}  \exp(\frac{\tau}{2 D_a}v^2) 
 \frac{d^{\nu}}{d v^{\nu}} \exp(-\frac{\tau}{2 D_a}v^2) .
 \end{equation}
By substituting the expansion in the  FPE~ \eqref{fpeq} when $k=0$ we obtain the following recursion relation for the amplitudes $n_\nu$:
\begin{equation}
\frac{\partial n_{\nu-2}(x)}{\partial x} +\nu \frac{D_a}{\tau} \frac{\partial n_\nu(x)}{\partial x}   
 = -\frac{\nu-1}{\tau}  n_{\nu-1}(x)
 + D_t \frac{\partial^2 }{\partial x^2} n_{\nu-1}(x) ,
 \label{hierarchy}
\end{equation}
with the condition $n_{-1}=0$ and $n_0(x)=\int dv f_c(x,v)$.
 We can, now, define
 the steady-state average polarization $m(x)$ as the first velocity moment of the distribution function:
 \be
 m(x)=\int dv v f_c(x,v) =\frac{D_a}{\tau} n_1(x) .
 \label{mn0}
 \ee
  Under stationary conditions and  $D_t=0$,  
 $m(x)$ is proportional to the local average of the active force (being $u=v$ in this region) 
 and thus vanishes in the absence of  external fields in virtue of Eq. \eqref{dotx}; this is seen by considering
 Eq.~\eqref{hierarchy}  for $\nu=1$ together with \eqref{mn0} :
\be
 \frac{\partial m(x)}{\partial x} 
 = D_t \frac{\partial^2 }{\partial x^2} n_0(x) ,
 \label{dadt}
 \ee
 and assuming a zero current condition.
  On the other hand, if $D_t>0$,  the fact that the external force is zero in some region of space does not necessarily imply the 
 corresponding vanishing of $m(x)$. 
 To show that,  let us use again Eq.~\eqref{dadt} and consider a finite density gradient term, $D_t   \frac{\partial n_0(x)}{\partial x}$: 
 a non uniform density profile is now sufficient to induce a polarization even where the external force acting on the AOUP is locally zero.
 As we shall see in section \ref{Pressure}, 
such a coupling between standard diffusion and polar order, represented by the $n_0(x)$ and $m(x)$ terms, respectively, determines an effective force between 
inclusions immersed in active suspensions.

Since Eq.~\eqref{hierarchy} represents an infinite hierarchy of coupled differential equations for the coefficients $n_\nu$, 
we need to introduce a suitable truncation able to capture
the phenomenology discussed in the previous Sections. Our truncation procedure comes easily in
the Hermite-basis and consists in setting $n_\nu(x)=0$ for all $\nu\geq n_{max}$, i.e.
in assuming corrections around a Gaussian-like approximation. The simplest possibility is to set
$\nu_{max}=2$ (case A), which leads to the so called screening-approximation. Instead, by considering $\nu_{max}=4$ 
(case B), the resulting approximation is equivalent to the hydrodynamic treatment based on the
first three moments of the velocity distribution~\cite{huang2009introduction} together with the idea that the term $n_3(x)$
(analogous to the heat flux) can be eliminated in favor of the spatial gradient of $n_2(x)$, i.e. to the gradient
of a kinetic temperature.

Thus we write the following equations:
 \begin{equation}
 \frac{1}{\tau} (1+\frac{D_a}{D_t})  m(x)
 -D_t \frac{\partial^2 }{\partial x^2} m(x)+2 \frac{D_a^2}{\tau^2} \frac{\partial n_2(x)}{\partial x}=0   \, ,
 \label{mequation}
\ee
\be
\frac{\tau}{D_a} \frac{\partial m(x)}{\partial x}   +\frac{2}{\tau} n_2(x)- D_t \frac{\partial^2 }{\partial x^2} n_2(x)+ 
3 \frac{D_a}{\tau}\frac{\partial n_3(x)}{\partial x} =0
 \ee
 where we used eq. \eqref{dadt} to eliminate $n_0(x)$. 
 In case A the closure is $n_2(x)=0$, 
 while in case B, in analogy with the phenomenological procedure followed in hydrodynamic treatments,
 one assumes
 $n_3(x)=- \frac{\tau}{3}\frac{\partial n_2(x)}{\partial x}$.
 Both approximations predict exponential solutions
 and for $D_a/D_t$ small enough a typical length over which the moments vary scaling proportionally to
 $\sqrt{D_t \tau \, \frac{D_t}{D_a+D_t}}$. In the following, for the sake of simplicity, we shall report only results concerning the so called 
 screening approximation (case A) first employed in reference  \cite{yan2015force} in the framework of the 
 ABP model.  The solution  reads
 \begin{equation}
n_0(x)=(n_w-n_M)\,\frac{ \cosh(\frac{x-L}{\lambda})}{\cosh(\frac{L}{\lambda})}+ n_M ,
 \label{screeningapprox}  
\ee
where $n_w=n_0(0)=n_0(2L)$ and $n_M$ depends on the geometry of the problem. The "polarization field" is given by:
\begin{equation}
m(x) = m_w  \frac{ \sinh(\frac{x-L}{\lambda})}{\cosh(\frac{L}{\lambda})}= \frac{ D_t}{\lambda} (n_w-n_M)\,  \frac{ \sinh(\frac{x-L}{\lambda})}{\cosh(\frac{L}{\lambda})}
\ee
where
\begin{equation}
\lambda^2= D_t \tau \, \frac{D_t}{D_a+D_t}
\label{screenlambda}
\ee
 and $m_w$ is the polarization field at the wall.


Notice that if $D_t=0$ the polarization field vanishes.
 The comparison in Fig.\ref{fig:pxprediction}  between the numerical $n_0(x)$ and the analytic prediction
 displays a fair agreement if $D_t$ is not too small with respect to $D_a$. 
 
 
\begin{figure}[!t]
\centering
\includegraphics[width=1\linewidth,keepaspectratio]{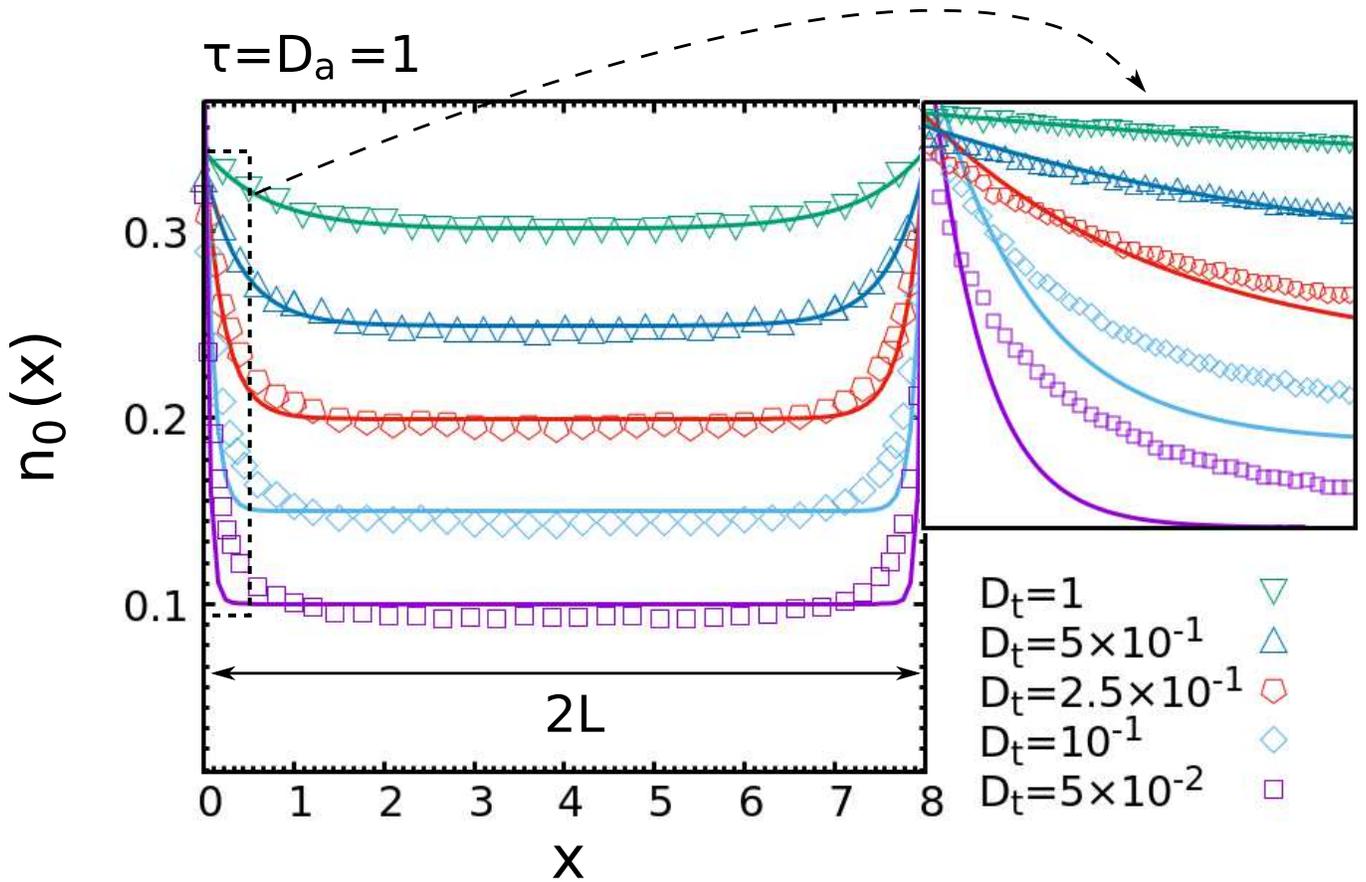}
\caption{System with translational noise for different values of $D_t$, as shown in the legend:
comparison between $n_0(x)$  computed numerically (symbols) and analytically from Eq.\eqref{screeningapprox} (solid lines)
where $\lambda$ is the screening length predicted by the Eq.~\eqref{screenlambda}.
Parameters: $D_a=\tau=1$, $k=10$.  For the sake of clarity
we applied to each curve a shift  $0.05$ with respect to the  curve below. In the inset we display a magnification of the region in the proximity of the wall.}
\label{fig:pxprediction}
\end{figure}

\begin{figure}[!t]
\centering
\includegraphics[width=1\linewidth,keepaspectratio]{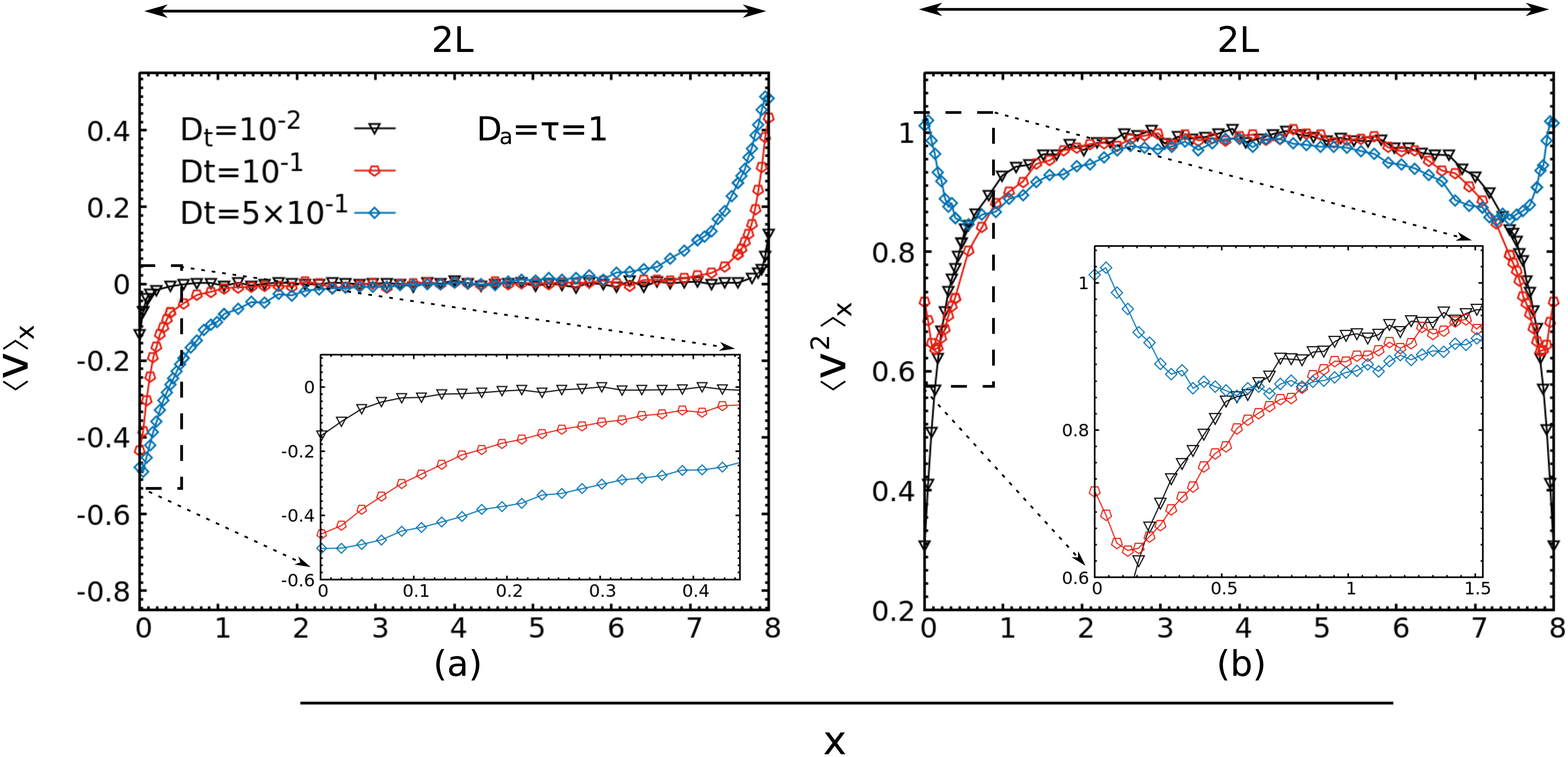}
\caption{
Panel (a) shows the average value of the first moment of the velocity $\langle v \rangle_x$ and 
the right panel its variance $\langle v^2 \rangle_x$ for $\tau=D_a=1$, $k=10$ and different values of $D_t$, as shown in the legend.}\label{fig:momentdt}
\end{figure}

 This result is consistent with the one obtained in ref.~\cite{yan2015force} for the ABP model, confirming that AOUP is a useful approximation of ABP which captures all the physical aspect of the accumulation near the walls also in the presence of thermal noise. In addition, our study sheds some light on the closure employed in the hydrodynamic-like approach, by considering the description in terms of the particle velocity.

As a consequence of the simultaneous presence of two baths (active and thermal), $\langle  v\rangle_x$ is non-zero
as shown in Fig.~\ref{fig:momentdt} (c) for different values of $D_t$: in a region close to the wall, $\langle  v\rangle_x\neq0$ and decays monotonically to 0. The decay length decreases as $D_t$ decreases until it disappears when $D_t \ll D_a$ and the thermal noise becomes negligible. 
The role of  thermal noise is not trivial producing a non-monotonic behavior: in a thin space region, close to the wall, $\langle v^2 \rangle_x$ decreases reaching a minimum and  then increases until it reaches the limit value $D_a/\tau$, as shown in the Fig.\ref{fig:momentdt}(d).
Such an effect can be accounted for theoretically by going beyond the Gaussian closure $n_2=0$, i.e. by truncating the coupled system of equations \eqref{hierarchy} at a higher level, but for space reasons we do not include this possibility in the present analysis.
\subsubsection{Beyond the Gaussian approximation for systems without thermal noise.} 

Indeed, in the absence of translational noise the screening approximation
Eq.~\eqref{screeningapprox} cannot be used because  the limit $D_t\to 0$ is singular.
 We also find numerically that in this limit $\langle  v\rangle_x=0$ as expected. In Fig.~\ref{fig:momentdt0}(a) we observe  that for $\tau=1$ and  $D_a=0.1$ and $1$
the second velocity moment $\langle v^2 \rangle_x$  grows until reaches the constant value $D_a/\tau$, the quadratic velocity moment of a uniform system.
 We can distinguish two different behaviors in the interval $[0,2L]$: a persistent region where $\langle v^2 \rangle_x\neq D_a/\tau$  where the influence of the wall remains important, and a far region characterized by $\langle v^2 \rangle_x\approx D_a/\tau$ which basically is bulk-like. Only in the case $D_a=10$ the curve does not saturate and there is no separation between the two regions, since the
 persistence length $\sqrt{D_a\tau}$ is comparable with $L$. 

Let us go back to the theory and see that, when $D_t=0$ and $\nu=2$, Eq.~\eqref{hierarchy}
predicts that, being $n_1(x)= n_0(x) \langle v \rangle_x/(D_a/\tau)=0$, the profile $n_0(x)$ is simply related to $n_2(x)$ by
$$
\frac{\partial n_{0}(x)}{\partial x} +2 \frac{D_a}{\tau} \frac{\partial n_2(x)}{\partial x} =0.
$$
 Hence,  the Gaussian approximation $n_2(x)=0$, which was employed to derive to formula~\eqref{screeningapprox}, fails when $D_t=0$ because 
 it would predict a constant profile $n_0$ in the force-free region: this is consistent with the UCNA approximation, but not with the numerical result.
The breakdown of the Gaussian approximation can be further ascertained by 
 checking the numerical data against the following relations which hold for a Gaussian distribution of velocities: $\langle v^{2n+1} \rangle_x=0$ and $\langle v^{2n} \rangle_x= C_{2n, 2} \langle v^2\rangle_x$, being $C_{2n, 2}$ the binomial coefficient.
The numerical study of the third and fourth moments of the velocity distribution, for a system with $D_t=0$, 
 displays evident deviations from the Gaussian predictions as illustrated in Fig.~\ref{fig:momentdt0}(b)
 where a non zero third moment of the velocity is reported and Fig.~\ref{fig:momentdt0}(c) where the distribution has a non-vanishing kurtosis.
 Having established that the Gaussian closure, $n_2(x)=0$, is unfit to capture the observed behavior, a possible remedy
to such a deficiency could be a higher order truncation of the series \eqref{hermiteexpansion}, taking into  account  terms $n_\nu(x)$ with $\nu=4$ in the Hermite expansion.
	Such a procedure leads to a solution for the space-density resembling Eq.~\eqref{screeningapprox} with a screening length,
$\lambda=\sqrt{5 D_a \tau/6}$,
proportional to the persistence lenght and capturing the phenomenology of the case $D_t=0$. 
This possibility is briefly discussed in Appendix  \ref{Gaussianapprox}, for space reasons, 
whereas here we report the  following approximate factorization of the fifth moment of the velocity 
 in terms of lower moments that we found  empirically  from our numerica data:
\begin{equation}
 \langle v^5\rangle_x \sim 10 \langle v^2 \rangle_x \langle v^3 \rangle_x ,
 \label{v5factor}
 \ee
  where the factor $10$ is a combinatorial factor which takes into account the number of ways
 of factorizing the average, is correct. As shown in
 Fig.~\ref{fig:momentdt0}(d) the comparison between
 the numerical estimates of $ \langle v^5\rangle_x$ and $10 \langle v^2 \rangle_x \langle v^3 \rangle_x $.
 corroborates the validity of the hypothesis expressed by Eq.~\eqref{v5factor}. In appendix \ref{Gaussianapprox}, we present an argument
 supporting this factorization of the average.
 

\begin{figure}[!t]
\centering
\includegraphics[width=1\linewidth,keepaspectratio]{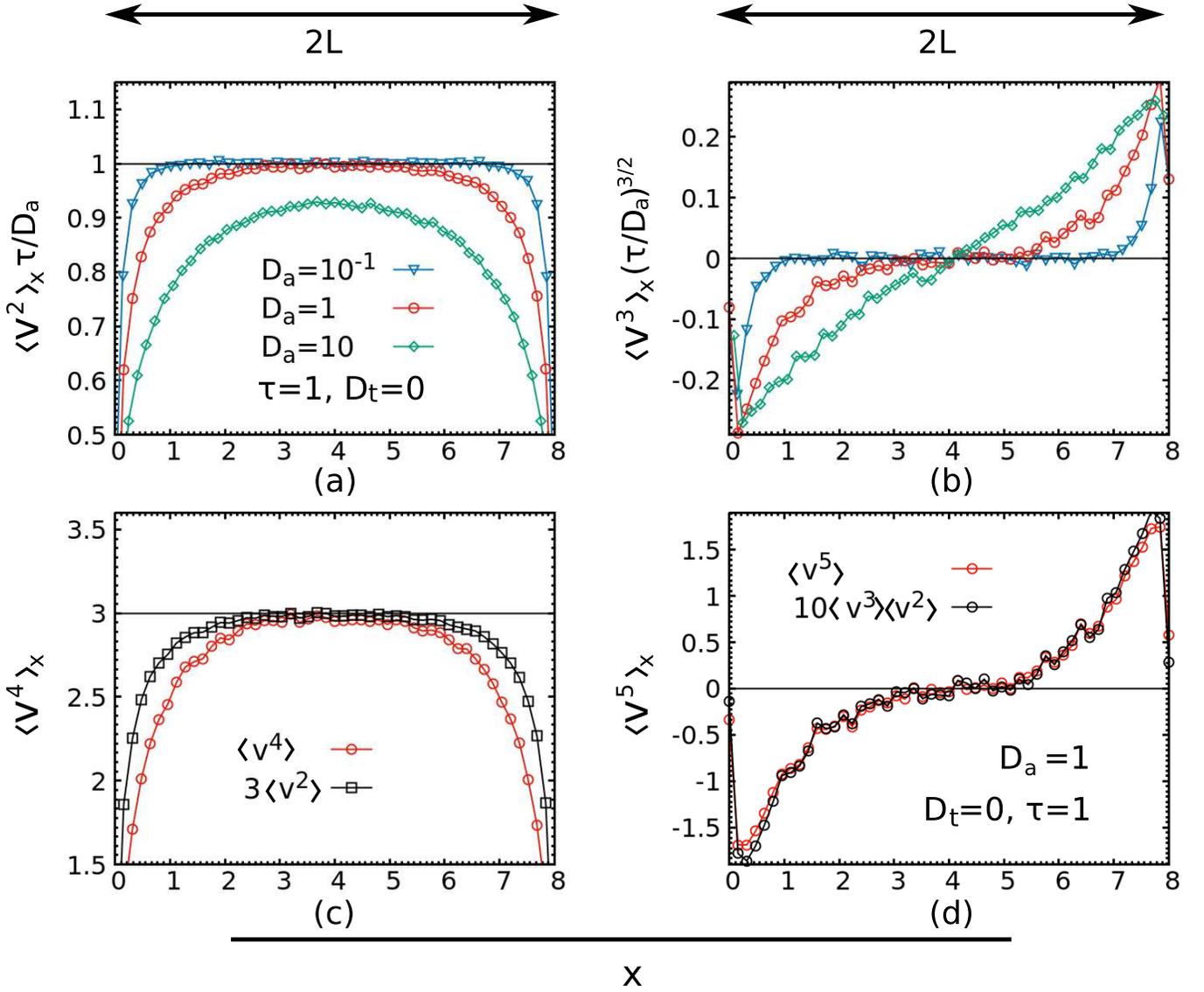}
\caption{Panel (a) and (b): the second and third moment of the
velocity $\langle v^2 \rangle_x$ and $\langle v^3 \rangle_x$ 
as a function of the position $x$ for $D_t=0$ and different values of $\sqrt{D_a \tau}$, as shown in the legend, and $k=10$.
In panel (c)  we display the comparison between $\langle v^4 \rangle_x$ and the Gaussian closure $3\langle v^2 \rangle_x$ when $\tau=D_a=1$ and $D_t=0$:  the deviation from the Gaussian prediction  $\langle v^3 \rangle_x=0$ and  $\langle v^4 \rangle_x=3\langle v^2 \rangle_x \langle v^2 \rangle_x$ is evident. 
In panel (d) in the case $\tau=D_a=1$ and $D_t=0$,
we compare the average $\langle v^5\rangle_x$ with the empirical closure $10 \langle v^2 \rangle_x \langle v^3 \rangle_x$. }
\label{fig:momentdt0}
\end{figure}

\section{Forces on the confining walls}
\label{Pressure}
We turn, now, to consider the mechanical properties of the confined active system and derive a formula for
the mechanical pressure, $\mathcal{P}_{wall}$, exerted on a harmonic wall by the active gas. To achieve that, we use the following  equation expressing the mechanical balance condition between the pressure exerted by the particles on the wall and the one exerted by the wall on the $N$ particles:
\begin{equation}
\mathcal{P}_{wall} = -N\int_{-\infty}^{0} dx \, n_0(x) \phi'(x) ,
\label{wallformula}
\ee
where the upper limit of the integral takes into account the fact that the left wall potential vanishes for any $x>0$.
Now, we compute $\mathcal{P}_{wall}$ at the left wall by substituting $\phi'(x)=kx$ and obtain:
\begin{equation}
\mathcal{P}_{wall} = N n_w D_a\gamma(\frac{1}{\Gamma}+\Delta) s(\Gamma,\Delta) ,
\label{wallpressure}
\end{equation}
where
$N n_w$ represents the numerical density at the wall $x=0, 2L$, the factor $D_a \gamma (\frac{1}{\Gamma}+\Delta)$ has the dimensions of a temperature and is the effective temperature of an active particle confined in a harmonic trap
\cite{szamel2014self,das2018confined}.
The last factor contained in the pressure formula~\eqref{wallpressure}
\begin{equation}
\begin{aligned}
s(\Gamma,\Delta)=
\frac{1}{1+\frac{\Gamma-1}{ 1+\Gamma\Delta}\left[ 1+\sqrt{1+\frac{ 1+\Gamma\Delta}{\Gamma-1}} \right] } \leq 1 ,
\end{aligned}
\end{equation}
is the result of two effects which can be observed when $\tau$ decreases: a) the shift of the peak of the density distribution
towards more negative values of $x$  and b) its broadening.
For $\tau=0$, being $\Gamma=s(\Gamma,\Delta)=1$,  Eq.~\eqref{wallpressure} reduces to $N n_w (D_a \gamma +D_t \gamma )$ 
which is the pressure of a suspension of Brownian particles against a wall, so that \eqref{wallpressure}  can be seen as the generalization to the active case
of the ideal gas formula.


\subsection{The wall boundary conditions}

In the pressure formula \eqref{wallpressure}, the constant $n_w$ is yet undetermined, and as we shall see below 
it can be fixed by specifying the system set-up, i.e. its geometrical and physical properties.
By multiplying the FPE~\eqref{fpeq}  by $v$ and integrating with respect to $v$ and using the no particle flux condition
$$
 \frac{D_a}{\tau} n_1(x)=D_t \, \frac{\partial} {\partial x}  n_0(x) ,
$$
 we obtain the following expression relating a total derivative to the wall force:
\begin{flalign}
& \frac{\partial} {\partial x} \left(  \left(\frac{D_a}{\tau} +\frac{D_t}{\tau}\right) n_0(x)+ 2  \frac{D_a^2}{\tau^2} n_2(x)  -D_t^2 \, \frac{\partial^2} {\partial x^2}  n_0(x)- \frac{D_t}{\gamma}  \phi''(x) n_0(x)\right)\nonumber\\
& = -\frac{\phi'(x)}{\tau \gamma}     \,n_0(x) .
 \label{stresscondition}
\end{flalign}
We now integrate with respect to $x$  Eq.~\eqref{stresscondition} between $-\infty$ and $\bar x$ 
 where  $\bar x>0$:  
\begin{flalign}
& \tau \gamma \, \Bigl( \frac{(D_a+ D_t)}{\tau} n_0(x) +2   \frac{D_a^2}{\tau^2} n_2 (x)  -D_t^2 \, \frac{\partial^2} {\partial x^2}  n_0(x)- \frac{D_t k}{\gamma}\theta(-x) n_0(x)  \Bigr)_{-\infty}^{\bar x}\nonumber\\
& =\frac{\mathcal{P}_{wall}}{N} ,
\label{pressurebalance}
\end{flalign}
where we have taken into account  Eq.~\eqref{wallformula}.

 \subsubsection{Semi-infinite system.}
 
  Now, we restrict our analysis to the case $D_t$ not too small with respect to $D_a$, thus excluding the singular limit $D_t=0$.
	In order to simplify the analysis, we consider the limit $\bar{x} \gg \lambda$, which allows us to assume that the term $n_2(\bar{x}) \approx 0$ and we evaluate the left-hand side of Eq.~\eqref{pressurebalance} using Eq.~\eqref{screeningapprox} with the
result:
\begin{equation}
\mathcal{P}_{wall}= D_a\gamma \,  \Bigl( 1+\Delta\Bigr) N n_M .
\label{pwsemiinfinite}
\ee
Using the explicit representation of $\mathcal{P}_{wall}$, Eq.\eqref{wallpressure}, we express
the probability density at the wall, $n_w$, in terms of  the probability density $n_M$ and of the parameters of the model:
\begin{equation}
n_w=n_M \, \frac{1+\Delta }{\frac{1}{\Gamma}+\Delta }\frac{1}{s(\Gamma,\Delta)}.
\label{nwsemiinfinite}
\ee
Let us remark that we always have $n_w>n_M$ because both factors $\frac{1+\Delta }{\frac{1}{\Gamma}+\Delta }
$ and $\frac{1}{s(\Gamma,\Delta)}$ are larger than $1$ if $\tau>0$, so that the wall density is higher than the density at midpoint
and we may argue that there is a positive surface excess. Only in the Brownian limit $\tau=0$ we obtain $n_w=n_M$
for all values of $k$.
If now we take the limit of a semi-infinite system $L\to \infty$, 
we have $n_0(x)=(n_w-n_M)e^{-x/\lambda}+ n_M$
and we can make the identifications
$\rho=N n_M$  and $\rho_w=N n_w$, where $\rho$ and $\rho_w$  are the bulk and wall numerical densities, respectively.
Finally, using the condition that $\mathcal{P}_{bulk}=\mathcal{P}_{wall}$, necessary in order to have mechanical equilibrium,
we identify the r.h.s. \eqref{pwsemiinfinite}  with the bulk pressure, $\mathcal{P}_{bulk}$, of a uniform system at density $\rho$, i.e. $\mathcal{P}_{bulk}=D_a\gamma \,  \Bigl( 1+\Delta\Bigr) \rho$.
\begin{figure}[!t]
\centering
\includegraphics[width=1\linewidth,keepaspectratio]{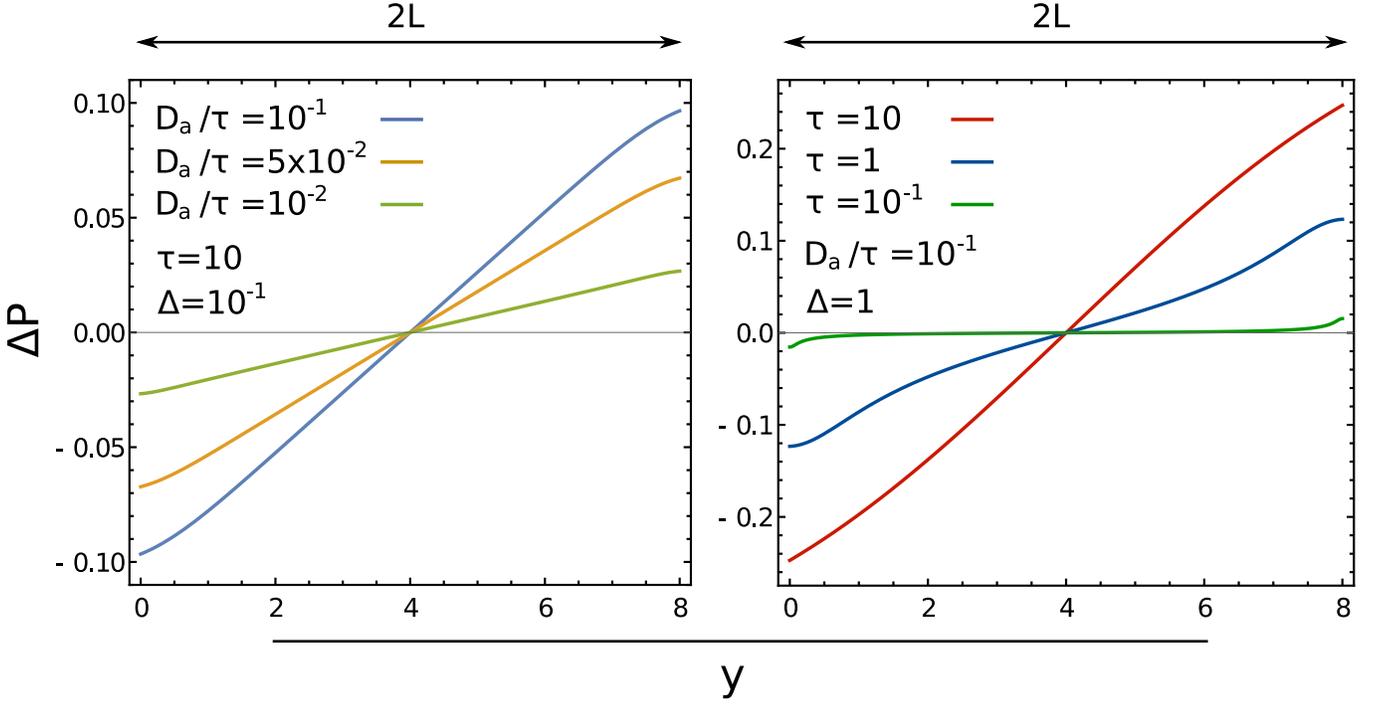}
\caption{Pressure difference between two compartments hosting two active suspensions having the same 
average numeric density of particles. Panel (a) corresponds to a system where the persistence time is fixed and so the ratio between the active and thermal diffusion coefficient, but the active power is varied. The largest active power corresponds to the largest pressure difference. 
 Panel (b): The active power is fixed, but the persistence time $\tau$ is varied. When $\tau$ increases the pressure difference increases
 as the system departs more and more from thermodynamic equilibrium.}
\label{fig:piston}
\end{figure}
\subsubsection{Wall Pressure in a slit system.} 
We turn, now, to study the pressure in a slit-like geometry with the purpose
of understanding how the force acting between two parallel plates immersed in a solution of active particles, 
depends on the  wall separation.
To find the density at the wall, $N n_w$ we must compute $n_0(x)$ in the whole space, match the expression \eqref{n0error} with \eqref{screeningapprox} at each wall and finally normalize the profile.
The probability density profile $n_0(x)$ can be written as:
\begin{equation}
n_0(x)=\theta(-x)n^{left}_0(x) + \theta(-x)\theta(2L-x) n_0^f(x) +\theta(2L-x)n^{right}_0(x) ,
\end{equation}
where $n^{left(right)}_0(x)$, the probability density distribution with  walls located at $0 \,(2L)$, is given by Eq. \eqref{n0error}
and $n_0^f(x)$ the density in the free region $[0,2L]$ is given by Eq.~\eqref{screeningapprox}.
From the normalization of the probability distribution we have the condition:
\begin{equation}
\int_0^{2L} dx n_0^f(x) + 2 \int_{-\infty}^{0} n^{left}_0(x) = 1 ,
\label{n0norm}
\ee
where the factor $2$ in the last term takes into account the symmetry of the two walls.
After performing the integrals and eliminating $n_M$ we obtain the relation:
\begin{equation}
\begin{aligned}
\frac{1}{n_w}&=\frac{2}{\sqrt \pi}\left(\frac{2\gamma D_a (1+\Gamma \Delta)}{k\Gamma}\right)^{1/2}\arctan \left( \sqrt{\frac{ (1+\Gamma \Delta)}{\Gamma-1}} \right) 
\\
&+2\lambda\tanh (\frac{L}{\lambda}) +2\frac{ \frac{1}{\Gamma} +\Delta} {1+\Delta} 
\frac{L-\lambda \tanh(\frac{L}{\lambda}) }
{1+\frac{\Gamma-1}{ 1+\Gamma\Delta}\left[ 1+\sqrt{1+\frac{ 1+\Gamma\Delta}{\Gamma-1}} \right] } .
\label{nwinv}
\end{aligned}
\end{equation}
We point out that the first term takes into account the finite width of the peak of $n_0(x)$ in the regions $x\leq0$ and $x\geq 2L$ due to the softness of the walls:  it vanishes for all values of the remaining parameters  when $k\rightarrow \infty$.
As a check we consider the equilibrium limit $\tau\to 0$
\begin{equation}
\begin{aligned}
&\lim_{\tau\to 0}\mathcal{P}_{wall} = Nn_wD_a \gamma\left(1+\Delta\right)\\
&\lim_{\tau\to 0} \frac{1}{n_w}= \left(\frac{2  \pi D_a\gamma (1+\Delta)}{k}\right)^{1/2}+2L \geq 2L
\label{inequal}
\end{aligned}
\end{equation}
as expected, since it corresponds to the situation of an overdamped passive system in contact with two independent white noise sources.
In the hard wall limit $k\to \infty$ and $\tau=0$  we find $n_w=1/2L$ and $\mathcal{P}_{wall}=\frac{ N}{2L} D_a \gamma\left (1+\Delta\right)$, the ideal gas equation of state corresponding to a system subjected to two white noise baths.

\subsection{Effective force between plates and Casimir effect}

Under the same hard-wall limit $k\to \infty$, but with $\tau>0$  one can see that there is an accumulation effect
at the wall: in fact, according to Eq.\eqref{nwinv} we have
 $n_w\geq 1/2L$.
Finally, the wall pressure can be computed inserting Eq.~\eqref{nwinv}
in Eq. \eqref{wallpressure}. For the sake of simplicity, we write
its expression in the limit $k\to \infty$ but $\tau$ finite:
\begin{equation}
\mathcal{P}_{wall}= D_a \gamma \frac{N}{2 L} \frac{ \left(1+\Delta\right)}{1+
\frac{\lambda}{L}  \tanh(\frac{L} {\lambda}) \left(\frac{1}{\Delta}+ \frac{(1+\Delta)^{3/2}}{\Delta^2}  \right)  } .
\label{pressurevsL}
\ee

In order to ascertain the activity induced force 
acting on parallel plates, we must compare the pressures of two systems having different sizes, $L_1$ and $L_2$, with $L_1<L_2$ but the same
average density of each system, $\bar \rho=\frac{N_i}{2 L_i}$, is identical. Clearly, the system
with the smaller size according to Eq.~\eqref{pressurevsL} will exert the smaller pressure on the walls.  We may conclude that
two parallel plates surrounded by a sea of
active particles and separated by a distance $2L$ will experience an effective attraction 
according to Eq. \eqref{pressurevsL}. Precisely, if $L_2\to \infty$, but $\bar \rho$ is fixed and $L_1/\lambda\gg 1$ we have
\begin{equation}
\mathcal{P}_{wall}^{(1)}-\mathcal{P}_{wall}^{(2)} \approx - D_a \gamma \bar \rho  \left(\frac{ \left(1+\Delta\right) }{\Delta}+ \frac{(1+\Delta)^{5/2}}{\Delta^2}  \right) 
\, \frac{\lambda}{L} .
\ee
Thus, in the low-density regime we consider we find 
 that the force increases linearly the active power $D_a$ and depends
 monotonically on plate separation  in agreement with the simulation results by Ni et al.   \cite{ni2015tunable} and the theoretical prediction of Vella et al. \cite{vella2017fluctuation} of a  decay $\propto 1/L$.

Such an effect can also be illustrated by the following gedanken-experiment:
let us consider a finite system and insert a third hard wall C, identical to the first two, at an arbitrary position $y \in (0, 2L)$, in such a way that the average numerical densities in the two
resulting  compartments are equal: $N_l / y = N_r/(2L - y)$ with $N_l+N_r=N$. 
According to the present  theory  the pressure difference between the left and right compartment is given by:
\begin{equation}
\begin{aligned}
\Delta \mathcal{P} =&D_a\gamma (1+\Delta) \Bigl[\frac{N_l}{ y} 
\frac{1}{1+\frac{\lambda}{y}  \tanh(\frac{y} {\lambda}) \left(\frac{1}{\Delta}+ \frac{(1+\Delta)^{3/2}}{\Delta^2}  \right)}   \\
&
-\frac{N_r}{(2L-y)}  \frac{1}{1+\frac{\lambda}{2L-y}  \tanh(\frac{2L-y} {\lambda}) \left(\frac{1}{\Delta}+ \frac{(1+\Delta)^{3/2}}{\Delta^2}  \right)} \Bigr ] .
\end{aligned}
\end{equation}
Of course,
for a Brownian system, if the average densities in the left and in the right compartments are set to be equal for any choice of the wall position $y$,  i.e. if the condition $N_l = N y/2L=N_r = N(2L -y)/2L$ is satisfied, 
the pressure difference, $\Delta \mathcal{P}$  vanishes.

The physical reason of such a phenomenon is strictly related to the accumulation of active particles in front of a wall and clearly emerges in the Eq.~\eqref{nwinv}: increasing $L$ the constant $n_w$ grows, meaning that more particles push on the wall and exert a larger pressure.

On the other hand, our prediction suggests a completely different situation in the active case which we show in the Fig.~\ref{fig:piston}. In particular $\Delta \mathcal{P}>0$ if $y>L$ and  $\Delta \mathcal{P}<0$ if $y<L$: in fact, the particles in the small 
compartment exert a smaller pressure on the wall C than the one exerted by those in the larger compartment, in spite of the fact that the numerical densities are equal.
For small separations $2L$ the approximation $n_2(x)\approx 0$ ceases to be correct and we do not expect that the force 
obeys anymore the scaling $L^{-1}$, however there is some room for improvement, for instance, by employing higher order closure approximations,
such as including terms $n_3,n_4$ etc, but in the present study we do not pursue such a possibility.

\section{Conclusions}
\label{Conclusions}
To conclude, let us remark that the physical effect of activity is twofold: i) it determines a non-uniform density profile because active particles accumulate near the wall and, in the case of deformable boundaries, penetrate inside them; on the contrary,
 a system of non-interacting Brownian particles in the hard wall limit would not develop any
density gradient;  ii) the pressure exerted on the walls of a slit of width $2L$
by a system of average density $\bar \rho$ depends on the wall separation.
The second phenomenon is relevant when the confinement length becomes comparable with the persistence length.
Hereafter, we summarize the main achievements of the present work. 

1)  We have extended the study of the AOUP to the case of one one-dimensional non-interacting particles under confinement,
and, going beyond the UCNA approximation,  we do not integrate out the active noise but retain it as a "velocity" variable.
By considering a simple parametrization of the bounding potential, we have been able under reasonable approximations
to derive simple expressions for  the density profile
and polar field and eliminate some negative features of the UCNA solution, such as its jumps in correspondence
of discontinuities of the potential, and its failure to account for polar order near a boundary. 
Our treatment introduces a healing length which  produces  smoother density profiles which have been found in good agreement with the 
results of numerical simulations.

2) The moment method employed to approximate the non equilibrium distribution function $f(x,v)$ 
predicts an effective force between two plates immersed in an active suspension, similar to the force in the classical Casimir effect: indeed, the attraction between the plates is due to the active particles, which could represent active bacteria, while low-density
Brownian particles, i.e. colloidal particles, do not exert any appreciable force on the plates.
The possibility of generating and controlling the force between immersed objects, for instance by tuning the illumination of an active suspension or modifying its concentration and/or temperature, is quite interesting and offers an alternative to other techniques which instead require the chemical modification of the surfaces. Finally, the effective temperature of such suspensions, which determines the intensity of such a force, can be higher than the solvent temperature~\cite{rohwer2017transient}.

Future work will concern the extension of the theory to higher dimensions in order to treat 
active solution-mediated interactions  between inclusions of more general shape~\cite{singh2017soft} or moving pistons \cite{caprini2017fourier}. Including the interaction among the particles 
is also a challenge and we may expect that with increasing density the excluded volume effects could lead to an effective
repulsion for some values of the plate separation.

\section*{Conflicts of interest}
There are no conflicts of interest to declare.

\section*{Acknowledgements}

We thank Andrea Puglisi and  Angelo Vulpiani for illuminating discussions.


\appendix

\section{Stationary distribution for the harmonic oscillator in the $(x,v)$ variables}
We know the full stationary solution of \eqref{fpeq}  for the harmonic oscillator. It can be written as a double Gaussian with
a velocity distribution whose peak changes with $x$. This peak corresponds to an x-dependent velocity, $\langle v\rangle_x,$ which is also the mean
velocity at fixed $x$. 
\begin{equation}
f_h(x,v)={\mathcal N} \exp\left(-\frac{k x^2}{2 \, D_a \gamma}\frac{\Gamma}{1+\Gamma\Delta}\right) \times \exp\left(\frac{ \left(v-\langle v\rangle_x\right)^2}{2\sigma^2_v}  \right)
\label{statpxv}
\ee
with
\begin{equation}
\langle v\rangle_x=  -\frac{  \Gamma\Delta}{  1+  \Gamma\Delta} \,   \frac{ k}{\gamma}x, \qquad \sigma^2_v= - \frac{D_a}{\tau \Gamma} \left(\frac{ 1+  \Gamma^2 \Delta}{ 1+  \Gamma \Delta}\right)
\label{valormedio}
\ee
Let us remark that $f_h(x,v)$ due to the presence of a non-vanishing average velocity $\langle v\rangle_x=$ has the form of the distribution function of a system {\it in local but not global equilibrium},
in contrast with the case $\Delta=0$.


 \section{How to rationalize the non-Gaussian closure}
 \label{Gaussianapprox}
 As we can see from the structure of the solution in the potential free-region the screening length vanishes when $D_t\tau \to 0$
 even though $D_a$ remains fixed. In order to remove such a nonphysical feature, we must consider carefully
 the limit $D_t=0$. In this case, the study hierarchy \eqref{hierarchy} becomes relatively simple and allows to predict
 a non-vanishing decay length and sheds some light on the form of the closure  \eqref{v5factor}.
 We begin by writing explicitly the hierarchy assuming $n_1(x)=0$ in this limit:
\bea
 &&
 \frac{\partial n_0(x)}{\partial x} +2 \frac{D_a}{\tau} \frac{\partial n_2(x)}{\partial x}   
 = 0 
\\&&
 3  \frac{D_a}{\tau} \frac{\partial n_3(x)}{\partial x}=    -\frac{2}{\tau} n_2(x)
 \\&&
 \frac{\partial n_2(x)}{\partial x} +4  \frac{D_a}{\tau}\frac{\partial n_4(x)}{\partial x}=    -\frac{3}{\tau} n_3(x)
 \\&&
 \frac{\partial n_3(x)}{\partial x} +5  \frac{D_a}{\tau}\frac{\partial n_5(x)}{\partial x} =    -\frac{4}{\tau} n_4(x)
\eea
An option is to break
 the hierarchy by setting $n_5(x)=0$ and after eliminating $n_4(x)$ we write
 \bea
&&
\frac{2}{\tau} n_2(x) +3 v_a^2 \frac{\partial n_3(x)}{\partial x}=0
\\&&
 \frac{\partial n_2(x)}{\partial x} +\frac{3}{\tau} n_3(x)-v_a^2 \tau \frac{\partial^2 n_3(x)}{\partial x^2}=0
 \eea
 We obtain a  closed set of linear differential equations that can be solved by combinations
 of exponentials of the form $e^{\pm \mu x}$, with $\mu$ determined by a simple algebraic equation. We find  
 $\mu^2=\frac{6}{5}\, \frac{1}{D_a \tau}$ and the profile is given by
 $n_0(x)=A\cosh(\mu x)+C$. We now try to verify the working hypothesis \eqref{v5factor}. Using the Hermite expansion \eqref{hermiteexpansion} when $n_1=0$ and $v_a^2=\frac{D_a}{\tau}$ we obtain the relations:
 \bea
 &&
n_0(x)  \langle v^2 \rangle_x=2 v_a^4 n_2(x)+ v_a^2 n_0(x)\\
 &&
 n_0(x)  \langle v^3\rangle_x=6 v_a^6 n_3(x)\\
 &&
 n_0(x)  \langle  v^4 \rangle_x=24 v_a^8 n_4(x)+12 v_a^6 n_2(x) +3 v_a^4 n_0(x) \\
 &&
 n_0(x)  \langle v^5 \rangle_x=60 v_a^8 n_3(x) +120 v_a^{10} n_5(x).
 \label{n5system}
 \eea
  
 Since we have assumed $n_5=$ we obtain the equality:
\begin{equation}
   \langle v^5 \rangle_x=10 v_a^2  \langle v^3\rangle_x
   \label{v5analytic}
   \ee
Such a relation is compatible with Eq.\eqref{v5factor} only in the regime when 
$\langle v^2\rangle_x$ can be
replaced the constant factor $v_a^2$ in Eq~\eqref{v5analytic}. This is possible if
the space dependent average $ \langle v^2\rangle_x-v_a^2=2 v_a^4 \frac{n_2(x)}{n_0(x)}\approx 0$, i.e. in a regime of small $\tau$.

The empirical relation  \eqref{v5factor} instead is consistent with the choice $n_5(x)=\frac{n_3(x) n_2(x)}{n_0(x)}$. However, the substitution of such a relation into Eqs.~\eqref{n5system} leads to a closed set of non-linear equations which cannot be solved by simple analytic methods.

\bibliographystyle{rsc} 
\providecommand*{\mcitethebibliography}{\thebibliography}
\csname @ifundefined\endcsname{endmcitethebibliography}
{\let\endmcitethebibliography\endthebibliography}{}

\end{document}